\documentclass[
  aps,
  prl,
  reprint,
  superscriptaddress,
  amsmath,
  amssymb
]{revtex4-2}

\usepackage{graphicx}
\usepackage{dcolumn}
\usepackage{bm}
\usepackage[colorlinks=true,allcolors=blue]{hyperref}
\begin{document}

\preprint{APS/123-QED}

\title{Self-organized hybrid scenario mediated by fishbone modes in the EAST tokamak}

\author{J.S. Geng}
%\thanks{These authors contributed equally to this work.}
\affiliation{Institute of Plasma Physics, Hefei Institutes of Physical Science, Chinese Academy of Sciences, Hefei, 230031, China}

\author{Y.Q. Chu}
%\thanks{These authors contributed equally to this work.}
\affiliation{Institute of Plasma Physics, Hefei Institutes of Physical Science, Chinese Academy of Sciences, Hefei, 230031, China}

\author{B.F. Gao}
\affiliation{Key Laboratory of Computing Power Network and Information Security, Ministry of Education, Shandong Computer Science Center (National Supercomputer Center in Jinan), Qilu University of Technology (Shandong Academy of Sciences), Jinan, 250103, China}

\author{L.Q. Xu}
\email{Contact author: lqxu@ipp.ac.cn}
\affiliation{Institute of Plasma Physics, Hefei Institutes of Physical Science, Chinese Academy of Sciences, Hefei, 230031, China}

\author{F. Chen}
\affiliation{Institute of Plasma Physics, Hefei Institutes of Physical Science, Chinese Academy of Sciences, Hefei, 230031, China}

\author{Z.Y. Qiu}
\affiliation{Institute of Plasma Physics, Hefei Institutes of Physical Science, Chinese Academy of Sciences, Hefei, 230031, China}

\author{J.P. Qian}
\affiliation{Institute of Plasma Physics, Hefei Institutes of Physical Science, Chinese Academy of Sciences, Hefei, 230031, China}

\author{X.Z. Gong}
\affiliation{Institute of Plasma Physics, Hefei Institutes of Physical Science, Chinese Academy of Sciences, Hefei, 230031, China}

\author{Q. Long}
\affiliation{Institute of Plasma Physics, Hefei Institutes of Physical Science, Chinese Academy of Sciences, Hefei, 230031, China}
\affiliation{University of Science and Technology of China, Hefei, 230026, China}

\author{Y.X. Li}
\affiliation{Institute of Plasma Physics, Hefei Institutes of Physical Science, Chinese Academy of Sciences, Hefei, 230031, China}
\affiliation{University of Science and Technology of China, Hefei, 230026, China}

\author{C.B. Wu}
\affiliation{Institute of Plasma Physics, Hefei Institutes of Physical Science, Chinese Academy of Sciences, Hefei, 230031, China}

\author{X.H. Wu}
\affiliation{Institute of Plasma Physics, Hefei Institutes of Physical Science, Chinese Academy of Sciences, Hefei, 230031, China}
\affiliation{University of Science and Technology of China, Hefei, 230026, China}

\author{Y.F. He}
\affiliation{Institute of Plasma Physics, Hefei Institutes of Physical Science, Chinese Academy of Sciences, Hefei, 230031, China}

\author{Y.C. Hu}
\affiliation{Institute of Plasma Physics, Hefei Institutes of Physical Science, Chinese Academy of Sciences, Hefei, 230031, China}

\author{N. Sun}
\affiliation{Institute of Plasma Physics, Hefei Institutes of Physical Science, Chinese Academy of Sciences, Hefei, 230031, China}

\author{P.J. Sun}
\affiliation{Institute of Plasma Physics, Hefei Institutes of Physical Science, Chinese Academy of Sciences, Hefei, 230031, China}

\author{Y.H. Zhuge}
\affiliation{Institute of Plasma Physics, Hefei Institutes of Physical Science, Chinese Academy of Sciences, Hefei, 230031, China}
\affiliation{University of Science and Technology of China, Hefei, 230026, China}

\author{W.M. Zhang}
\affiliation{Institute of Plasma Physics, Hefei Institutes of Physical Science, Chinese Academy of Sciences, Hefei, 230031, China}
\affiliation{National Institute for Fusion Science, 322-6 Oroshi-cho, Toki, 509--5292, Japan}

\author{L. Zhang}
\affiliation{Institute of Plasma Physics, Hefei Institutes of Physical Science, Chinese Academy of Sciences, Hefei, 230031, China}

\author{W.B. Liu}
\affiliation{Institute of Plasma Physics, Hefei Institutes of Physical Science, Chinese Academy of Sciences, Hefei, 230031, China}

\author{H.L. Zhao}
\affiliation{Institute of Plasma Physics, Hefei Institutes of Physical Science, Chinese Academy of Sciences, Hefei, 230031, China}

\author{Y.F. Jin}
\affiliation{Institute of Plasma Physics, Hefei Institutes of Physical Science, Chinese Academy of Sciences, Hefei, 230031, China}

\author{S.X. Wang}
\affiliation{Institute of Plasma Physics, Hefei Institutes of Physical Science, Chinese Academy of Sciences, Hefei, 230031, China}

\author{K.N. Geng}
\affiliation{Institute of Plasma Physics, Hefei Institutes of Physical Science, Chinese Academy of Sciences, Hefei, 230031, China}

\author{K. Jia}
\affiliation{Institute of Plasma Physics, Hefei Institutes of Physical Science, Chinese Academy of Sciences, Hefei, 230031, China}
\affiliation{University of Science and Technology of China, Hefei, 230026, China}

\author{Y.X. Jie}
\affiliation{Institute of Plasma Physics, Hefei Institutes of Physical Science, Chinese Academy of Sciences, Hefei, 230031, China}

\author{H. Lian}
\affiliation{Institute of Plasma Physics, Hefei Institutes of Physical Science, Chinese Academy of Sciences, Hefei, 230031, China}

\author{R.J. Zhu}
\affiliation{Institute of Plasma Physics, Hefei Institutes of Physical Science, Chinese Academy of Sciences, Hefei, 230031, China}
\affiliation{University of Science and Technology of China, Hefei, 230026, China}

\author{S.Z. Yuan}
\affiliation{Institute of Plasma Physics, Hefei Institutes of Physical Science, Chinese Academy of Sciences, Hefei, 230031, China}
\affiliation{University of Science and Technology of China, Hefei, 230026, China}

\author{S.B. Zhang}
\affiliation{Institute of Plasma Physics, Hefei Institutes of Physical Science, Chinese Academy of Sciences, Hefei, 230031, China}

\author{H.Q. Liu}
\affiliation{Institute of Plasma Physics, Hefei Institutes of Physical Science, Chinese Academy of Sciences, Hefei, 230031, China}

\author{Q. Zang}
\affiliation{Institute of Plasma Physics, Hefei Institutes of Physical Science, Chinese Academy of Sciences, Hefei, 230031, China}

\author{Y.D. Li}
\affiliation{Institute of Plasma Physics, Hefei Institutes of Physical Science, Chinese Academy of Sciences, Hefei, 230031, China}

\author{P. Li}
\email {Contact author: lipan@ipp.ac.cn}
\affiliation{Institute of Plasma Physics, Hefei Institutes of Physical Science, Chinese Academy of Sciences, Hefei, 230031, China}

\date{\today}

\begin{abstract}
Self-organized hybrid scenario mediated by fishbone modes in the Experimental Advanced Superconducting Tokamak (EAST) is reported in this work. In this regime, internal transport barriers (ITBs) and a flat core safety factor profile were maintained simultaneously by fishbone modes. First,  the drive of low-frequency flows by fishbone modes is clarified, which subsequently suppresses turbulence transport. Besides, fishbone-induced poloidally symmetric (\(m=0\)) toroidal current sheets were first reported experimentally, which drive outward current redistribution that mitigates core current peaking. Together, these effects enable hybrid operation with stable ITBs and robust magnetohydrodynamic (MHD) stability, revealing a self-organized pathway for sustaining hybrid scenario for fusion reactors.
\end{abstract}

\maketitle

%\noindent
Magnetically confined nuclear fusion holds great significance for the realization of clean energy. The hybrid scenario offers improved confinement and higher \(\beta_{\mathrm{N}}\) compared to conventional H-modes\cite{bib1}, and a promising candidate for fusion reactors to achieve long-pulse high-fusion-gain plasma operation \cite{bib2,bib3}.

In the hybrid scenario, the central safety factor \(q_0\) close to unity, with a broad region of weak magnetic shear \cite{bib1,bib4}. Sustaining internal transport barriers (ITBs) is essential for maintaining steep pressure gradients, high energy confinement, and substantial bootstrap-current generation for hybrid scenario. The onset of the ITBs is controlled by multivariable coupling, whereas maintaining the ITB is challenged by the stability of self-organized structures, core impurity accumulation, and MHD instabilities \cite{bib5,bib6,bib7}. Another major difficulty is to maintain the weakly-shear q-profile for long pulse durations in hybrid scenario. Slow drift of qmin below unity triggers large-amplitude sawtooth oscillations and may render the hybrid scenario ineffective\cite{bib4,bib8}.

\begin{figure*}[t]
  \centering
  \includegraphics[width=\textwidth]{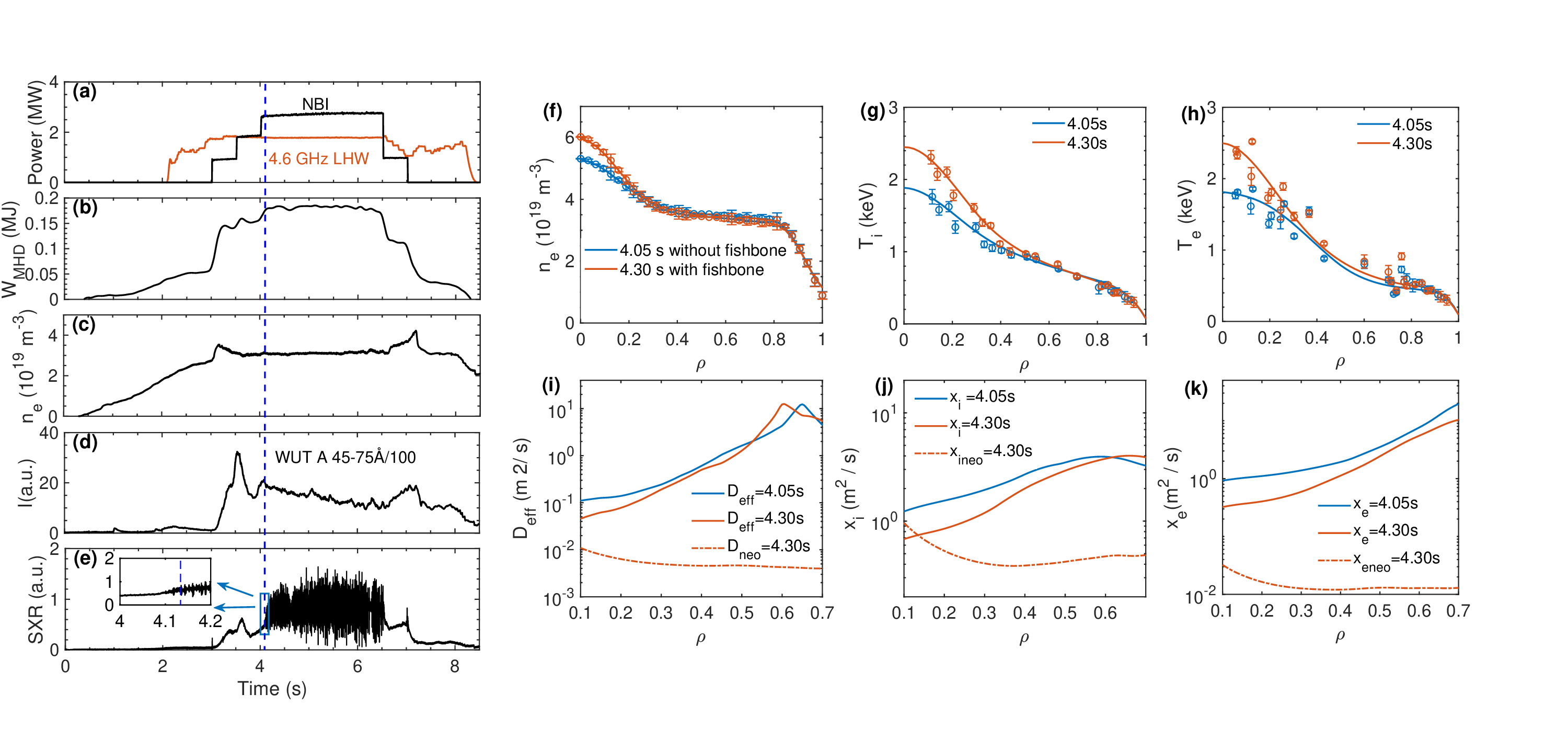}
  \caption{ (a) Absorbed Neutral-beam-injection power and \(4.6~\mathrm{GHz}\) lower-hybrid-wave heating power; (b) plasma stored energy; (c) line-averaged electron density; (d) tungsten emission measured by extreme ultraviolet spectrometer in the wavelength range of \(45\)--\(75~\AA{}\) and (e) the core soft X-ray emission. The inset in e highlights the onset of fishbone oscillations, and the vertical blue dashed line indicates the fishbone modes onset at \(t = 4.14~\mathrm{s}\). The profiles of (f) plasma density, (g) ion temperature, and (h) electron temperature before fishbone modes onset (\(t = 4.05~\mathrm{s}\), blue) and during fishbone activity (\(t = 4.30~\mathrm{s}\), orange). The solid curves are fits to the measurements. Corresponding radial profiles of (i) particle diffusivity, (j) ion thermal diffusivity, and (k) electron thermal diffusivity. The dashed curves denote the corresponding neoclassical transport coefficients at \(4.30~\mathrm{s}\).}
  \label{fig:overview}
\end{figure*}

Energetic particles provide an additional and potentially beneficial degree of freedom in plasma self-organization. The fishbone mode \cite{bib9,bib10} nonlinearly interacts with background drift-wave turbulence via three-wave resonance,modifying zonal flows. Transient or even sustained improvement in main plasma confinement is observed following fishbone bursts on JET \cite{bib11}, DIII-D \cite{bib12}, ASDEX Upgrade \cite{bib13}, EAST \cite{bib14}, and HL-2A \cite{bib15}. Gyrokinetic simulations of DIII-D tokamak showed that fishbone-driven zonal flows can reduce the energetic-particle loss and suppress turbulence by flow shear effect \cite{bib16}. Besides, fishbone activity may also contribute to the current-profile regulation\cite{bib17}. Experiments on ASDEX Upgrade suggested that fishbone modes could redistribute core current to maintain the central safety factor \(q_0\) close to unity \cite{bib18}. 
Such redistribution could limit core current peaking and provide a link between energetic-particle-driven MHD activity and macroscopic equilibrium regulation. These two features of the fishbone mode have been validated by simulations, though comprehensive experimental evidence remains lacking. Moreover, they suggest potential utility for hybrid scenario studies, and this aspect remains to be elucidated.

Here, we combine experimental measurements with numerical modeling to investigate the role of fishbone modes in a high-confinement hybrid plasma. We show that fishbone activity is associated with the development of low-frequency flows (\(f<2\,\mathrm{kHz}\)) that suppress turbulence transport and sustain ITBs. We further identify \(m=0\) toroidal current sheets localized near the \(q=1\) surface and demonstrate that they mitigate core current peaking, maintain a broad safety-factor profile, and reduce the tendency toward sawtooth relaxation. These results reveal a coupled pathway in which fishbone modes regulate flow, transport, and current evolution, thereby enabling a self-organized hybrid regime with sustained ITBs and robust MHD stability. More broadly, our findings suggest that energetic-particle-driven instabilities can be incorporated into the design of high-performance reactor scenarios as active elements of plasma self-organization.

%\section{Results}
%\section{Stationary fishbone operation with improved core confinement}\label{sec1}
%\noindent

In the EAST tokamak, hybrid-scenario discharges accompanied by fishbone modes have been reproducibly obtained. A representative discharge is presented in Fig.~1, with a plasma current \(I_p = 0.5\ \mathrm{MA}\), a toroidal field \(B_T = 1.6\ \mathrm{T}\) (clockwise), and an edge safety factor \(q_{95} = 3.7\). It is characterized by a noninductive current fraction of approximately \(75\%\), a central safety factor close to unity, and weak magnetic shear, consistent with hybrid-scenario operation. A \(4.6\ \mathrm{GHz}\) lower hybrid wave (LHW) system and three neutral beam injection (NBI) sources are used to heat the plasma. As shown in Fig.~1a, LHW is applied during the current flat-top from \(2.0\ \mathrm{s}\) onward. The absorbed NBI power is \(0.9\ \mathrm{MW}\) at \(3.0\ \mathrm{s}\), and increases to \(2.7\ \mathrm{MW}\) at \(4.0\ \mathrm{s}\). As the NBI power increases, the energetic-particle fraction increases in the plasma, the fishbone modes were excited by the resonance of energetic particles at \(4.14\ \mathrm{s}\). Following fishbone modes onset, the plasma stored energy reaches \(\sim 0.18\ \mathrm{MJ}\) and subsequently remains stable (Fig.~1b). Simultaneously, the core tungsten concentration gradually decreases during the fishbone burst period (Fig.~1d). The internal transport barriers (ITBs) in electron density, ion temperature, and electron temperature form as fishbone modes emerge, as shown in Fig.~1f--1h. The effective transport coefficients (particle diffusivity \(D_{\mathrm{eff}}\), ion thermal diffusivity \(\chi_i\), and electron thermal diffusivity \(\chi_e\)) were calculated and revealed a decrease of approximately one order of magnitude in the \(\rho < 0.3\) region (Fig.~1i--1k), which verified the reduction of anomalous transport. The ITB-containing hybrid scenario was sustained during the fishbone modes phase and collapsed upon fishbone modes termination.
%\section{Low-frequency flow generation and turbulence suppression}\label{sec2}

In hybrid scenarios, magnetic shear remains weakly positive across the core, without strong reversed-shear region  \cite{bib19}. Therefore the \(E \times B\) flow shear required for ITB formation and sustainment is critical and highly sensitive to plasma state  \cite{bib14,bib20,bib21}. Stable operation of the ITB-containing hybrid scenario for a duration exceeding ten times the energy-confinement time was achieved in this study. The low-frequency (\(f < 2\ \mathrm{kHz}\)) component of the poloidal rotation velocity inferred from the Doppler shifts  \cite{bib22,bib23,bib24} was used to characterize the low-frequency flow in poloidal direction. As shown in Fig.~2, with the injection of the third NBI beam at \(t = 4.0\ \mathrm{s}\), the intensity of turbulence at the intermediate wavenumber (\(k_\theta = 12\ \mathrm{cm}^{-1}\), \(k_\theta \rho_s \sim 1.6\)--\(2.9\)) and the high wavenumber (\(k_\theta = 22\ \mathrm{cm}^{-1}\), \(k_\theta \rho_s \sim 3.1\)--\(4.8\)) measured by CTS system \cite{bib25,bib26,bib27} increased. However, the turbulence intensity trend reversed after the fishbone modes appeared. The turbulence intensity continuously decreased, while the low-frequency flow gradually increased until it peaked at approximately \(5.5\ \mathrm{s}\) and then remained stable, the ITB was stably maintained during this period. While fishbone modes persisted, the enhanced low-frequency flow suppressed turbulence and reduced plasma anomalous transport.

\begin{figure}[t]
  \centering
  \includegraphics[width=\columnwidth]{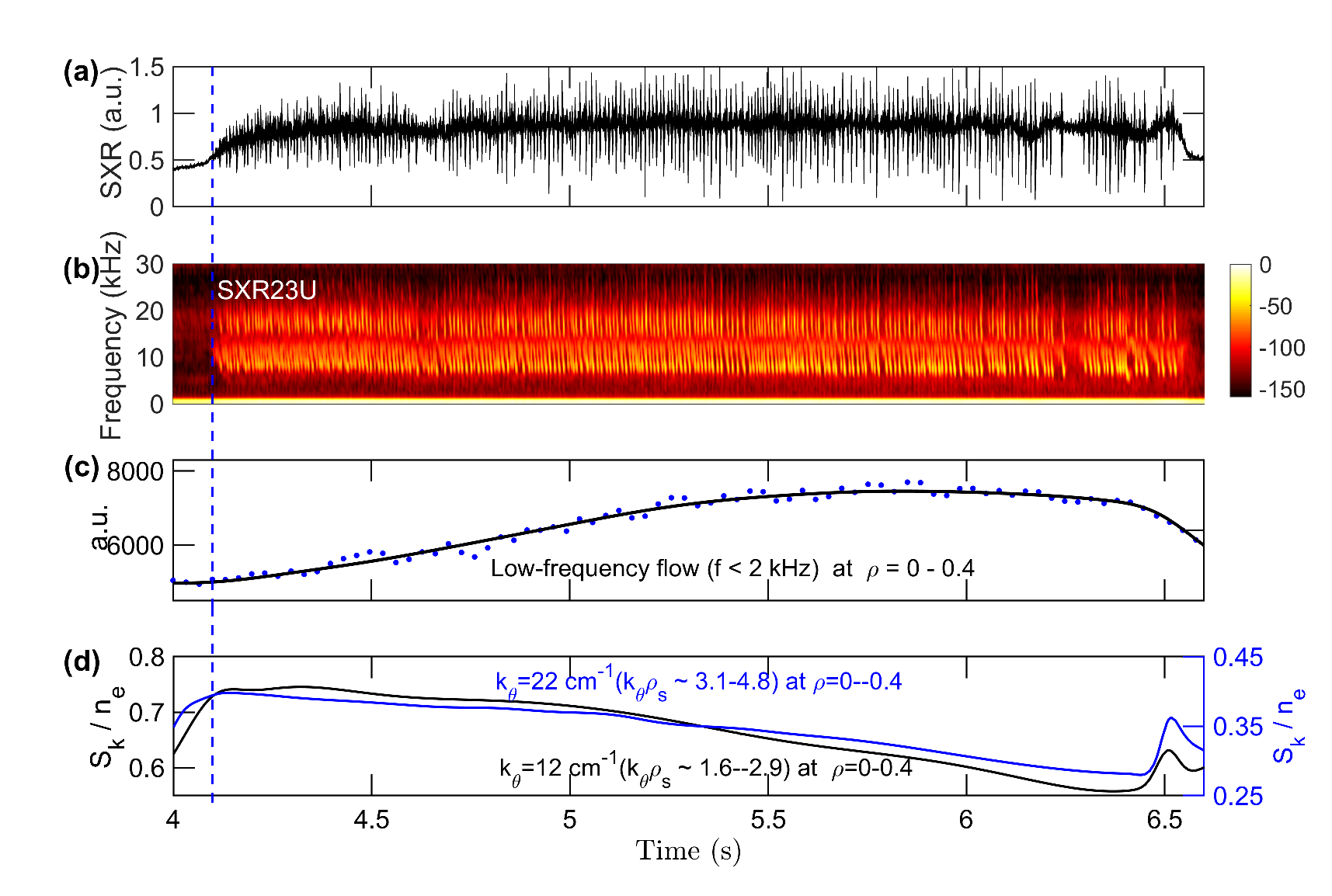}
 \caption{(a) Core soft-x-ray signal and (b) correspond frequency spectrum. The vertical blue dashed line marks the onset of the fishbone modes. (c) Low-frequency component (\(f < 2\ \mathrm{kHz}\)) of the poloidal rotation velocity inferred from Doppler shifts by the CTS system in the region \(\rho \sim 0\)--\(0.4\). The blue dots denote the CTS measurements, and the solid curve is a fit to the data. (d) Plasma core normalized turbulence amplitudes measured by the CTS system at intermediate (\(k_\theta = 12\ \mathrm{cm}^{-1}\), \(k_\theta \rho_s \sim 1.6\)--\(2.9\); black) and high (\(k_\theta = 22\ \mathrm{cm}^{-1}\), \(k_\theta \rho_s \sim 3.1\)--\(4.8\); blue) wavenumbers.}
\label{fig:flow}
\end{figure}
%\noindent

To identify the relationship between this low-frequency flow and the fishbone modes, the bicoherence spectrum for the envelope of magnetic-perturbation signals measured by Mirnov coils was calculated over the frequency range of \(0\)--\(50\ \mathrm{kHz}\) to identify the coupling characteristics of fishbone modes. The auto-bicoherence spectra of the magnetic perturbation signals before and after the fishbone bursts are shown in Fig.~3a and 3b, respectively. Before the onset of the fishbone modes (\(t=3.90\ \mathrm{s}\)), no significant coupling structure was observed in the low-frequency range. While a strong coupling band appears at \(f_1 + f_2 < 2\ \mathrm{kHz}\) after the fishbone modes appears (\(t=4.30\ \mathrm{s}\)), and the frequency range of this coupling band is consistent with the low-frequency flow. These results indicate nonlinear coupling between fishbone modes and the low-frequency fluctuations, supporting the interpretation that the observed low-frequency flow is driven by fishbone modes rather than the background electromagnetic fluctuations.

\begin{figure}[t]
  \centering
   \includegraphics[width=\columnwidth]{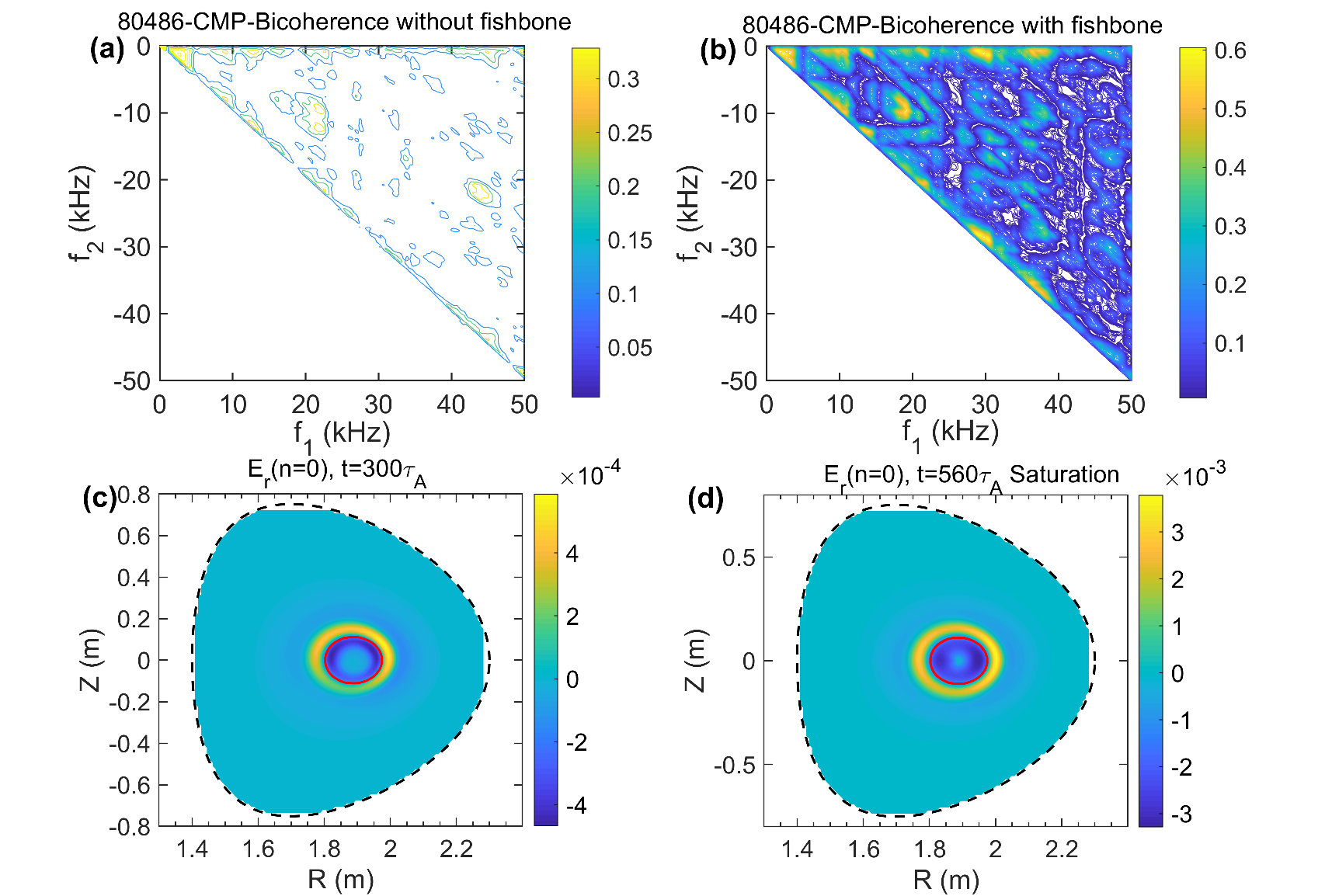}
  \caption{Auto-bicoherence analysis of Mirnov signals (a) before and (b) after the fishbone bursts. Spatial distribution of the \(n=0\) radial electric field amplitude \(|E_r|\) obtained from nonlinear M3D-K simulations during (c) the early nonlinear phase (\(t = 300\,\tau_A\)) and (d) the fishbone saturation phase (\(t = 560\,\tau_A\)). Note that the red solid line represents the \(q = 1\) rational surface.}
  \label{fig:bicoherence}
\end{figure}

To further investigate the feature of the fishbone-driven low-frequency flow, the nonlinear simulations of the resonance interaction between energetic particles and fishbone modes have been performed using the M3D-K code  \cite{bib28,bib29}. The simulations are based on the equilibrium parameters after fishbone modes appeared (\(t=4.30\ \mathrm{s}\)). Before fishbone modes saturation (Fig.~3c, approximately \(300\ \tau_A\), where \(\tau_A\) represents the Alfvén times), the simulated \(n=0\) component of \(E_r\) driven by fishbone modes were mainly localized around the \(q=1\) rational surface and remained relatively weak. As the fishbone modes evolve into the saturation phase (Fig.~3d, approximately \(560\ \tau_A\)), the \(n=0\) component of \(E_r\) near the \(q=1\) surface increased by one order of magnitude, and it exhibits a poloidal symmetry (\(m=0\)) feature. Fishbone modes nonlinearly excited a radial electric field characterized by toroidally symmetric (\(n=0\)) and poloidally symmetric (\(m=0\)), which contributed to the zonal shear flow with low-frequency characteristics.

The simulation implies that the observed fishbone-driven low-frequency flow might be the low-frequency zonal flow. The above results support the mechanism that fishbone modes suppress turbulence transport by driving plasma flow \cite{bib30,bib31,bib32}. The fishbone-driven zonal flows exhibited long-time stable characteristics and satisfied the requirements for ITB sustainment in hybrid operating scenario.

%\section{Radial outward redistribution of current density and resilient current profile}\label{sec3}

\begin{figure}[t]
\centering
\includegraphics[width=\columnwidth]{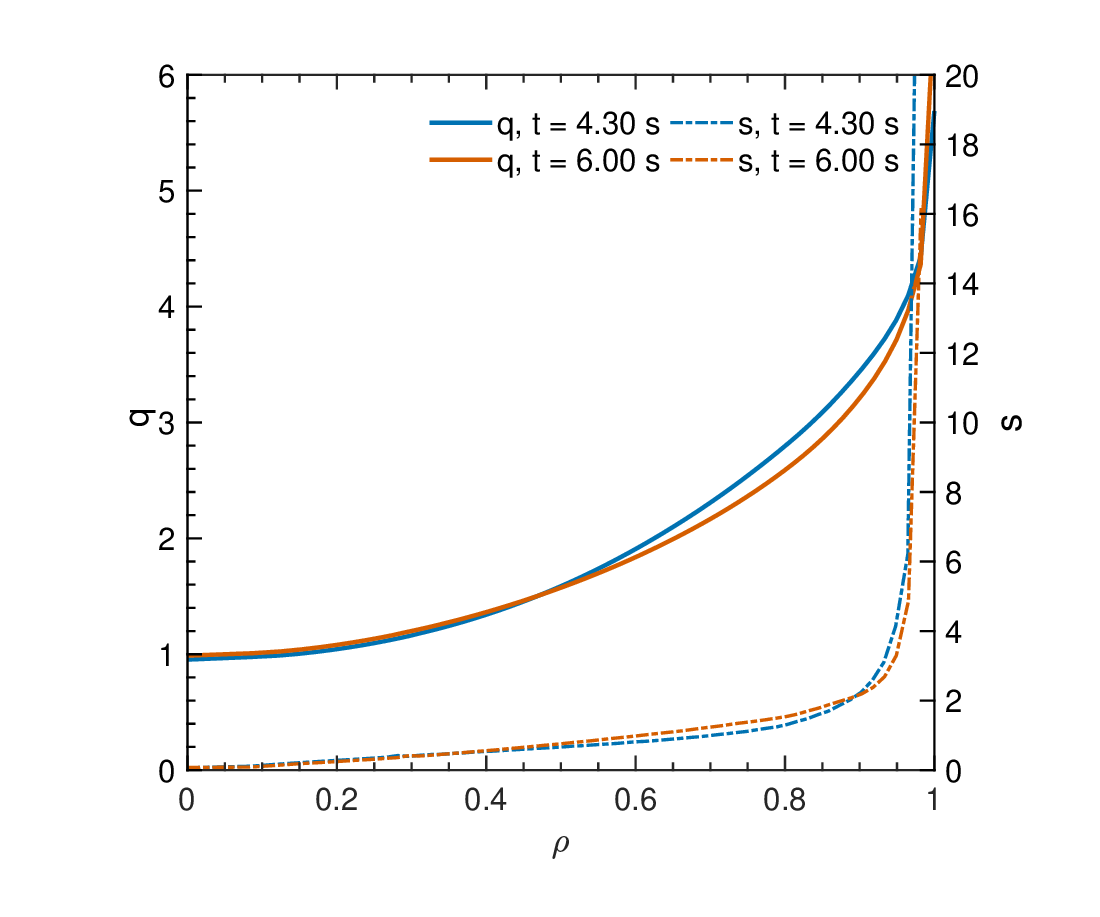}
\caption{Radial profiles of the safety factor \(q\) (solid curves, left axis) and magnetic shear \(s\) (dashed curves, right axis) at \(t = 4.30\text{ s}\)(blue)  and \(t = 6.00\text{ s}\)(orange) during sustained fishbone activity. The central \(q\) profile remains close to unity with weak magnetic shear in the core region.}
\label{fig:qprofile}
\end{figure}

%\noindent
Beyond the challenge of sustaining internal transport barriers, another major difficulty is to maintain the weakly-shear core \(q\)-profile for long pulse durations in hybrid scenario \cite{bib19}. Slow drift of \(q_{\min}\) below unity triggers large-amplitude sawtooth oscillations and may render the hybrid scenario ineffective. Stable control of the current profile was accomplished throughout the hybrid operating campaign, featuring \(q_{\min} \sim 1\) and weak magnetic shear in the plasma core (seen in Fig.~4). To investigate the effect of fishbone modes on the current density profile, the line-integrated density and Faraday rotation angle of discharge \#80486 were measured based on the high temporal resolution polarimeter-interferometer (POINT) diagnostic \cite{bib33,bib34}, where \(\Phi = 2.82 \times 10^{-15} \lambda n_e dl\) and \(\psi = 2.62 \times 10^{-13} \lambda^2 n_e B_\parallel dl\). As can be inferred from the formula, for each measurement chord, the Faraday rotation angle generally varies with the density. In addition, the sign of the Faraday rotation angle depends exclusively on the direction of \(B_\parallel\), where \(B_\parallel\) is the magnetic-field component along the laser propagation direction. For the POINT sightlines, the measured field component (\(B_\parallel\)) is dominated by the radial projection of the poloidal magnetic field that generated by the toroidal plasma current [47]. Notably, at this time, the \(Z\)-position of the magnetic axis lies between the 5th (\(Z = 0.085\ \mathrm{m}\)) and 6th (\(Z = 0\ \mathrm{m}\)) chords (see Fig.~5c). 

\begin{figure}[t]
\centering
\includegraphics[width=\columnwidth]{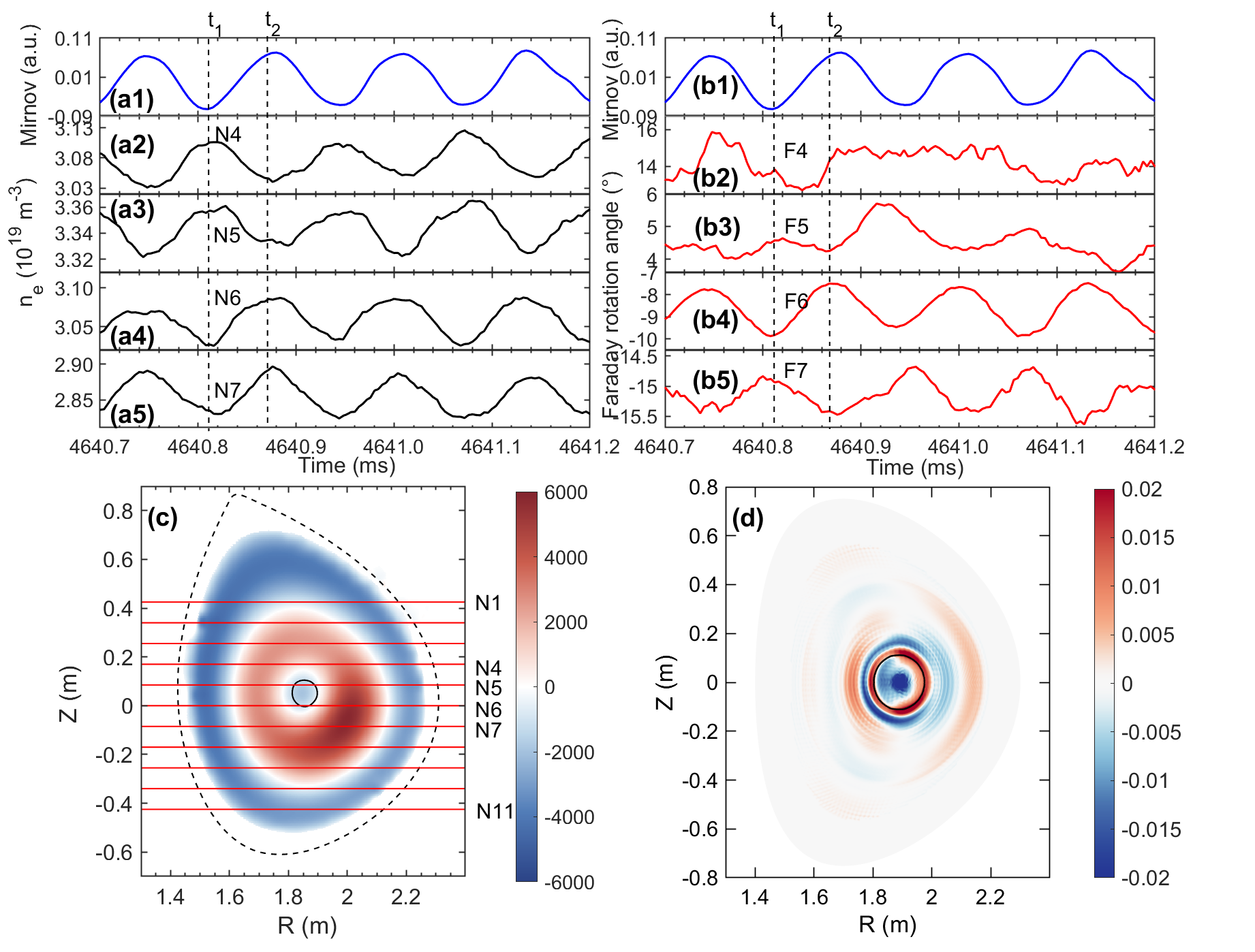}
\caption{(a) Mirnov-coil signal and electron-density responses measured along POINT chords N4--N7 during a single fishbone mode cycle. (b) Corresponding Faraday-rotation-angle signals along chords F4--F7. Vertical dashed lines mark the fishbone mode trough (\(t_1\)) and peak (\(t_2\)). (c) Reconstructed toroidal current-density perturbation, \(\Delta J = J(t_2) - J(t_1)\), in the poloidal cross-section. Red and blue denote positive and negative perturbations, respectively; the white contour marks \(\Delta J = 0\), the black circle indicates the \(q = 1\) surface, and the red horizontal lines show the POINT sightlines. (d) Toroidal current density perturbation with \(n=0\) from nonlinear M3D-K simulations, \(\Delta J_{n=0} = J_{n=0}(\mathrm{saturation}) - J_{n=0}(\mathrm{excitation})\).}
\label{fig:current}
\end{figure}

As the fishbone-induced magnetic perturbation increases from its trough at \(t_1 = 4640.81\ \mathrm{ms}\) to its peak at \(t_2 = 4640.87\ \mathrm{ms}\), the line-integrated densities on 5th and 6th chords exhibited an anti-phase density variation, the density on the 5th chord decreased while that on the 6th chord increased. This can be attributed to the fact that fishbone modes possesses mode numbers \(m/n = 1/1\). At the same time, the absolute value of the Faraday rotation angle decreased on the 5th and 6th chords but increased on the 7th chord. These observations suggest the following hypothesis: the core current is expelled from the \(q<1\) surface outward, leading to a redistribution of the poloidal magnetic field. Similarly, from the peak (\(t_2\)) to the trough of the fishbone modes magnetic perturbation signal, the absolute value of the Faraday rotation angle increased on the 5th and 6th chords but decreased on the 7th chord, which suggests that the core current increases within the \(q=1\) surface and decreases outside the \(q=1\) surface.

To identify the the spatial feature of current density perturbation induced by fishbone modes, the poloidal cross section of the toroidal current density perturbation, \(\Delta J = J(t_2) - J(t_1)\), is calculated by Faraday rotation angle and EFIT equilibrium, where \(t_1\) and \(t_2\) correspond to the trough and peak of the fishbone modes amplitude, respectively. As shown in Fig.~5c, the color scale represents \(\Delta J\): positive values (red) denote an increase in current density, whereas negative values (blue) denote a decrease. The white contour indicates \(\Delta J = 0\), which is close to the \(q=1\) rational surface (black circle). As the fishbone modes amplitude increases from the trough to the peak, the current density \(J\) decreases inside the \(q=1\) rational surface and increases outside the \(q=1\) rational surface, indicating that the current inside the \(q=1\) surface is expelled outward. The maximum current-density perturbation expelled from the \(q=1\) surface to the outside is approximately \(6.0\times 10^3\ \mathrm{A/m^2}\), which is only approximately \(1\%\) of the local background current density (\(\sim 6.0\times 10^5\ \mathrm{A/m^2}\)). Notably, the current perturbation is poloidally symmetric, demonstrating that the fishbone mode excites a current perturbation with poloidal symmetry (\(m = 0\)).

This interpretation is further supported by the \(n=0\) toroidal current density perturbation obtained from nonlinear M3D-K simulations (Fig.~5d), defined as \(\Delta J_{n=0} = J_{n=0}(\mathrm{saturation}) - J_{n=0}(\mathrm{excitation})\). As the fishbone mode evolves from the excitation phase to saturation, the perturbation is negative inside the \(q=1\) rational surface but positive outside the \(q=1\) rational surface, indicating an outward redistribution of current from inside the \(q=1\) surface. The maximum outwardly redistributed current perturbation expelled from the \(q=1\) surface to the outside is approximately \(2\%\) of the local background current density, which is comparable in order of magnitude to the \(\sim 1\%\) perturbation obtained experimentally in Fig.~5c. The agreement in both the radial structure and the perturbation magnitude supports the interpretation that fishbone modes drive the experimentally observed \(m=0\) current perturbation.

In plasma discharge with NBI heating, energetic particles and beam-driven current are deposited mainly in the plasma core \cite{bib35}. In the absence of an effective relaxation channel, the current profile can become overly peaked, eventually favoring sawtooth activity or other core MHD relaxation events, which often limits further confinement improvement and may decrease the stability of high-performance plasma \cite{bib4,bib36}. Fishbone activity may provide an alternative, relatively benign relaxation channel for the core current profile. Earlier ASDEX Upgrade experiments proposed that fishbone bursts can periodically redistribute the core current and thereby maintain a flat 
central \(q\) profile \cite{bib18,bib31,bib37,bib38}. In the present work, fishbone modes drive an outward redistribution of current from inside the \(q = 1\) surface, thereby preventing the continuous peaking of the core current. Recurrent fishbone bursts continuously regulate the local current profile, which helps to maintain the central safety factor near unity and sustain weak magnetic shear in plasma core. The hybrid scenario with fishbone modes characterized by weakly positive central magnetic shear with \(q_{\min} \sim 1\), the \(q\)-profile effectively avoided sawtooth oscillations.
  
%\section{Discussion and Summary}

This study achieved a stationary hybrid scenario featuring fishbone modes and uncovered the underlying mechanism for the compatible sustainment of ITBs and current profiles. Firstly, the increase of plasma core low-frequency flows driven by fishbone modes suppresses plasma core turbulence and reduces the anomalous transport, leading to the maintenance of ITBs. Second, a fishbone-induced perturbed toroidal current with poloidally symmetric (\(m=0\)) structure near the \(q = 1\) rational surface is first identified in experiments. The perturbed current structure drives radial outward redistribution of core current density, thereby establishing the resilient current density profile with \(q_{\min} \sim 1\) and weakly positive central magnetic shear.

Besides, weak magnetic shear may enhance the ability of flow shear to suppress turbulence \cite{bib39}. Ultimately, the fishbone-maintained current profile avoids sawtooth oscillations, strengthens turbulence stabilization by low-frequency flows, and enables hybrid operation with ITBs and robust MHD stability. Considering the rich energetic-particle population expected in burning plasmas, the hybrid scenario with fishbone modes constitutes an attractive candidate for future fusion reactors.

\begin{acknowledgments}
\textit{Acknowledgments}——We thank the EAST team for their support of the experiment, providing data, analysis, and related simulations presented in this article. This work was supported by the Strategic Priority Research Program of Chinese Academy of Sciences (grant nos. XDB0790101), the National MCF Energy R\&D Program (grant nos. 2024YFE03130004, 2022YFE03050000, 2024YFE03190002, 2024YFE03190004, 2025YFE03010000 and 2025YFE03200200) and the National Natural Science Foundation of China (grant nos.12305254, 12375229, 12375230, 12305253, 12005262 and 12275316).

J. S. G. and Y. Q. C. are co-first authors.

\textit{Data and materials availability:} The experimental data that support the findings of this study were obtained from EAST discharge \#80486. They are available from the corresponding authors upon reasonable request. The simulation data that support the outcome of this work are also available from the corresponding authors upon reasonable request.

\end{acknowledgments}

\end{document}